\documentclass[jgrga]{agutex}

\usepackage{graphicx}
\usepackage{amsmath}
\usepackage{amsfonts}
\usepackage{amssymb}
\usepackage{dsfont}

\newcommand{\be}{\begin{equation}}
\newcommand{\ee}{\end{equation}}
\newcommand{\bs}{\begin{subequations}}
\newcommand{\es}{\end{subequations}}

\newcommand{\ez}{\ensuremath{\hat{\boldsymbol{e}}_z}}

\newcommand{\pa}{\ensuremath{_\parallel}}

\newcommand{\Om}{\ensuremath{\varOmega}}

\newcommand{\Ga}{\ensuremath{\varGamma}}

\newcommand{\f}[1]{\ensuremath{\boldsymbol{#1}}}
\newcommand{\m}[1]{\ensuremath{\left\langle #1\right\rangle}}

\authorrunninghead{TAUTZ \& SHALCHI}

\titlerunninghead{PALMER CONSENSUS REVISITED}

\authoraddr{R. C. Tautz,
Zentrum f\"ur Astronomie und Astrophysik, Technische Universit\"at Berlin, Hardenbergstra\ss e 36, D-10623 Berlin, Germany.
(rct@gmx.eu)}

\authoraddr{A. Shalchi, Department of Physics and Astronomy, University of Manitoba, Winnipeg, Manitoba R3T~2N2, Canada.
(andreasm4@yahoo.com)}

\begin{document}

\title{Simulated Energetic Particle Transport in the Interplanetary Space: The Palmer Consensus Revisited}



\authors{R. C. Tautz \altaffilmark{1} and A. Shalchi, \altaffilmark{2}}

\altaffiltext{1}{Zentrum f\"ur Astronomie und Astrophysik, Technische Universit\"at Berlin, Hardenbergstra\ss e 36, D-10623 Berlin, Germany.}
\altaffiltext{2}{Department of Physics and Astronomy, University of Manitoba, Winnipeg, Manitoba R3T~2N2, Canada.}

\begin{abstract}
Reproducing measurements of the scattering mean free paths for energetic particles propagating through the solar system has been a major problem in space physics. The pioneering work of Bieber et al. [Astrophys. J. 420, 294 (1994)] provided a theoretical explanation of such observations, which, however, was based on assumptions such as the questionable hypothesis that quasi-linear theory is correct for parallel diffusion. By employing a hybrid plasma-wave/magnetostatic turbulence model, a test-particle code is used to investigate the scattering of energetic particles. The results show excellent agreement with solar wind observations.
\end{abstract}

\begin{article}

\section{Introduction}

The transport of energetic particles in a turbulent medium has been considered a key problem in various subfields of astrophysics. However, only in the solar system can one obtain detailed information about the transport parameters from measurements. Therefore, the solar wind provides the perfect laboratory to test our understanding of particle propagation through a tenuous plasma. In his pioneering paper, \citet{pal82:con} presented a detailed comparison of analytical particle diffusion coefficients along the mean magnetic field with real data. He found that the understanding of particle transport theory was quite incomplete at this time. The theoretical description for this comparison was based on quasi-linear theory combined with a simplified turbulence model (magnetostatic slab without turbulence dissipation range).

A few years later, \citet{bie94:pal} improved the theoretical description of particle transport by incorporating three important effects: (i) the one-dimensional slab model was replaced by a quasi three-dimensional model known as the slab/2D composite model. In the two-component model, the turbulence is described as a superposition of slab modes and so-called two-dimensional (2D) modes; (ii) dynamical turbulence effects were included to account for the fact that magnetic field vectors decorrelate with respect to the initial conditions as time passes; (iii) a strongly decreasing part of the turbulence spectrum was included, which is known as the dissipation range (see \citep{sch10:dis,che10:ani} and references therein). Therefore, a complete spectrum was used that contained energy, inertial, as well as dissipation ranges. As demonstrated by \citet{bie94:pal}, solar wind observations of energetic particles were successfully reproduced.

Since then our understanding of plasma-particle interactions has significantly improved (see, e.\,g., \citep{sha09:nli} for a review). It was shown that quasi-linear theory itself, which was the basis for \citep{bie94:pal}, is highly questionable \citep{tau06:wav} mainly due to: (i) the well-know \emph{90$^{\,\circ}$ scattering problem} describes the inability of the quasi-linear approach to describe correctly the pitch-angle scattering at pitch-angles close to $90^\circ$, which problem has been known since the early seventies of the 20\textsuperscript{th} century \citep{vol73:per} and has been revisited more recently \citep{tau08:soq,sha09:ana}; (ii) $90^\circ$ scattering has been explored in numerical work \citep{qin09:dmm} showing that quasi-linear theory is not reliable in general case; (iii) the so-called the \emph{geometry problem} is ascribed to the observation \citep{sha04:wnl,sha07:par} that quasi-linear theory overestimates the parallel mean free path for the standard slab/2D model used in \citep{bie94:pal}. All three problems are related to the applicability of quasi-linear theory to model the parallel transport of energetic particles in the solar system.

In this paper, in contrast, we revisit the Palmer consensus by using advanced numerical simulations. In test-particle codes, particle diffusion coefficients can be computed without employing a transport theory such as quasi-linear theory. Specifically, the so-called \textsc{Padian} code \citep{tau10:pad} will be used to compute the parallel diffusion coefficient for a hybrid turbulence model consisting of both plasma waves and magnetostatic turbulence and for the dissipative turbulence spectrum used in \citep{bie94:pal}.
In Sec.~2 we briefly discuss the turbulence model used in our simulations and in Sec.~3 we describe the test-particle code. In Sec.~4 we present the new numerical results as well as a comparison with observations. In Sec.~5 we conclude and summarize.

\section{The Turbulence Model}

The magnetic field scenario adopted in the present work can be approximated by a superposition of a homogeneous mean magnetic field and a stochastic component, i.\,e., by $\f{B} (\f{x}) = B_0 \f{e}_z + \delta \f{B} (\f{x})$. The first contribution corresponds to the solar magnetic field on a sufficiently small scale. A set of Cartesian coordinates is chosen so that the mean magnetic field is aligned with the $z$ direction. The second contribution is more difficult to model. The stochastic contribution is generated by the solar wind turbulence and is described in the following paragraph.

\subsection{The Magnetic Correlation Tensor}

The fundamental quantity in the theory of magnetic turbulence is the time-dependent correlation tensor in wave-vector space. The tensor is defined as
\be
\mathsf P_{\ell m} ( \f k,t)=\m{\delta B_\ell(\f k,t)\,\delta B_m^\ast(\f k,0)},
\ee
where we used the \emph{ensemble average} $\langle\dots\rangle$ and the wavevector $\f k$. A common assumption in turbulence and diffusion theory is that all tensor components have the same temporal behavior, i.\,e., $\mathsf P_{\ell m}(\f k,t)=\mathsf P_{\ell m}(\f k)\varGamma(\f k,t)$, where the magnetostatic tensor has \citep{mat81:str,rs:rays}, for axisymmetric turbulence, the form
\be
\mathsf P_{\ell m}(\f k)=G(k_\parallel,k_\perp)\left(\delta_{\ell m}-\frac{k_\ell k_m}{k^2}\right).
\ee
Furthermore, we introduced the so-called \emph{dynamical correlation function} $\varGamma(\f k,t)$, which is discussed below. The function $G(k_\parallel,k_\perp)$ describes the scale dependence of the magnetic fluctuation and is discussed in what follows.

\subsection{Two-component Turbulence}

In the solar system, the standard model for the magnetic field fluctuations is the so-called \emph{slab/2D composite} model. In this model, which is sometimes called the \emph{two-component model}, it is assumed that the spectrum can be approximated by a superposition of slab modes defined as $\delta \f{B} (\f{x}) = \delta \f{B} (z)$ as well as two-dimensional (2D) modes with $\delta \f{B} (\f{x}) = \delta \f{B} (x,y)$. Physically, the former modes represent parallel Alfv\'enic fluctuations and the latter modes idealize the quasi-2D structures including \emph{flux tubes} that can develop from the interactions of such waves \citep{bel71:alf,she83:ani}.

The two-component model can be confirmed by using extensive analyses of solar wind data \citep{mat90:mal,nar10:wav}. Such observations indicate that approximately $80\%$ of the turbulent inertial range energy budget is in the two-dimensional fluctuations \citep{bie96:two}. The rest of the magnetic energy is contained in the slab modes. More recent data analysis projects confirmed this structure \citep{hor08:ani,tur11:ani,tur12:mhd}. Furthermore, there is a strong support of the aforementioned turbulence model by numerical simulations \citep{mat96:3dt,sha07:cas,dmi09:mhd} and analytical treatments of turbulence \citep{zan93:mhd}. Additionally, the slab/2D model is very tractable in analytical studies of magnetic field lines and energetic particles \citep{mat95:tra,bie94:pal,sha10:ana}.

In the model described here, the turbulence power spectrum has the form
\be
G(k_{\parallel},k_{\perp}) = g^{\mathrm{slab}} (k_{\parallel})\,\frac{\delta (k_{\perp})}{k_{\perp}}
+ g^{\mathrm{2D}} (k_{\perp})\,\frac{\delta (k_{\parallel})}{k_{\perp}}
\ee
where the one-dimensional spectra are $g^{\mathrm{slab}}$ for slab modes and $g^{\mathrm{2D}}$ for two-dimensional modes. Furthermore, parallel and perpendicular wavenumber components are denoted by $k_{\parallel}$ and $k_{\perp}$, respectively, and $\delta (x)$ is the Dirac delta distribution.

\subsection{The Turbulence Spectra}

For the two turbulence spectra the form proposed by \citet{bie94:pal} is used, namely for the slab spectrum
\begin{align}
& g^{\mathrm{slab}} (k_{\parallel}) = \frac{C(s)}{2 \pi}\,\ell_{\mathrm{slab}} \delta B_{\mathrm{slab}}^2 \nonumber\\
\times &
\begin{cases}
\;{\displaystyle\left[ 1 + \left( k_{\parallel} \ell_{\mathrm{slab}} \right)^2 \right]^{-s/2}}, & \text{for } k_{\parallel} \leqslant k_{\mathrm{slab}}\\
\;{\displaystyle\left[ 1 + \left( k_{\mathrm{slab}} \ell_{\mathrm{slab}} \right)^2 \right]^{-s/2} \left( \frac{k_{\mathrm{slab}}}{k_{\parallel}} \right)^{p}}, & \text{for } k_{\parallel} > k_{\mathrm{slab}}
\end{cases}.
\label{specslab}
\end{align}
From the condition that the wavenumber-integrated correlation tensor be equal to the total magnetic field energy, the normalization function is found to be
\begin{equation*}
C(s)=\frac{1}{2\sqrt\pi}\,\frac{\Ga(s/2)}{\Ga((s-1)/2)}
\end{equation*}
with the gamma function $\Ga(z)$. In the spectrum used here, we introduce the slab bendover scale $\ell_{\mathrm{slab}}$, the slab dissipation wavenumber $k_{\mathrm{slab}}$, the inertial range spectral index $s$, and the dissipation range spectral index $p$. Furthermore we denote the total magnetic field strength of the slab modes as $\delta B_{\mathrm{slab}}$. For the two-dimensional spectrum, $g^{\mathrm{2D}}$, we employ precisely the same form, namely
\begin{align}
& g^{\mathrm{2D}} (k_{\perp}) = \frac{2 C(s)}{\pi}\,\ell_{\mathrm{2D}} \delta B_{\mathrm{2D}}^2 \nonumber\\
\times &
\begin{cases}
\;{\displaystyle\left[ 1 + \left( k_{\perp} \ell_{\mathrm{2D}} \right)^2 \right]^{-s/2}}, & \text{for } k_{\perp} \leqslant k_{\mathrm{2D}} \\
\;{\displaystyle\left[ 1 + \left( k_{\mathrm{2D}} \ell_{\mathrm{2D}} \right)^2 \right]^{-s/2} \left( \frac{k_{\mathrm{2D}}}{k_{\perp}} \right)^{p}}, & \text{for } k_{\perp} > k_{\mathrm{2D}}
\end{cases}.
\label{spec2D}
\end{align}
The parameters have the same physical meaning as those used in the slab spectrum, Eq.~\eqref{specslab}.

\begin{table}[t]
\caption{Summary of the various parameters assumed for the test-particle simulations.}

\vspace{1ex}
\renewcommand{\arraystretch}{1.2}
\begin{tabular}{lll}\hline\hline
Parameter	description & Symbol & Value\\
\hline
Inertial range spectral index 	& $s$ 						& $5/3$\\
Damping range spectral index		& $p$						& $3$\\
Alfv\'en speed					& $v_{\mathrm A}$				& $33.5$ km/s\\
Slab bend-over scale			& $\ell_{\mathrm{slab}}$ 		& $0.03$\,AU\\
2D bend-over scale 				& $\ell_{\mathrm{2D}}$			& $0.03$\,AU\\
Slab dissipation wavenumber		& $k_{\mathrm{slab}}$			& $3 \times 10^{6} \; (\mathrm{AU})^{-1}$\\
2D dissipation wavenumber		& $k_{\mathrm{2D}}$ 			& $3 \times 10^{6} \; (\mathrm{AU})^{-1}$\\
Mean magnetic field strength		& $B_0$						& $4.12$\,nT\\
Slab turbulence strength			& $\delta B_{\mathrm{slab}}^2$ 	& $0.2\:\delta B^2$\\
2D turbulence strength			& $\delta B_{2D}^2$				& $0.8\:\delta B^2$\\
\hline\hline
\end{tabular}
\label{ta:param}
\end{table}

Of course, there are alternative models (e.\,g., \citep{sri94:tur,gol95:tur}) that take into account the different physical origins of one- and two-dimensional fluctuations (cf. \citep{mat07:spe}). Here, however, identical spectra for both directions are assumed for reasons of (i) simplicity and (ii) to enable future comparisons with non-linear calculations that are only tractable analytically for as simple as possible spectral distributions.

\subsection{Dynamical Turbulence}

The \emph{dynamical correlation function} \citep{rs:rays,sha06:dyn} can be expressed as
\be
\Ga (\f {k}, t) = \cos \left( \omega t \right) e^{- \alpha t},
\ee
where the trigonometric and exponential factors describe oscillations related to wave-propagation effects and damping effects, respectively. Previously different models have been discussed to approximate the parameter $\alpha$. Examples are different plasma wave damping models \citep{ach93:dis,ach93:tra}, the \emph{damping model of dynamical turbulence} \citep{bie94:pal}, and the \emph{non-linear anisotropic model of dynamical turbulence} \citep{sha06:dyn}.

\begin{figure*}[tb]
\centering
\includegraphics[width=0.75\linewidth]{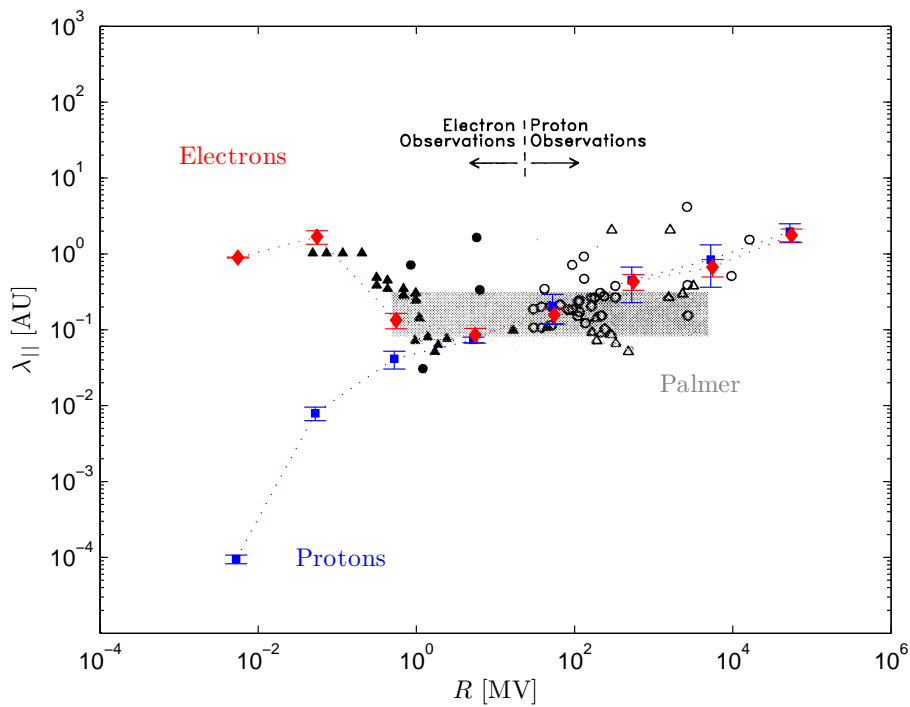}
\caption{(Color online) The parallel mean free path as a function of the particle rigidity for electrons (red diamonds) and for protons (blue squares) from the numerical simulations using the \textsc{Padian} code. The dissipation range spectral index is $p=3$ and the dissipation wavenumber is $k_d=3\times10^5\;(\mathrm{AU})^{-1}$. The comparison with the observational data (Palmer consensus range \citep{pal82:con}) shows excellent agreement. Data points obtained from Fig.~1 of \citet{bie94:pal}, reproduced by permission of the American Astronomical Society (AAS).}
\label{ab:palmerPEb}
\end{figure*}

Here, we choose $\alpha=0$ as test-particle codes are still unable to take into account damping effects. In contrast, the recently developed \textsc{Padian} code \citep{tau10:pad} can be applied for undamped plasma wave turbulence (see, e.\,g., \citep{tau10:wav}). Therefore, we use the simple model of undamped parallel propagating shear Alfv\'en waves corresponding to $\omega = j v_{\mathrm A} k_{\parallel}$ and $\alpha = 0$, where the Alfv\'en speed $v_{\mathrm A}=B_0/\sqrt{4\pi\rho}$ represents the speed of the propagating plasma wave. The parameter $j$ is used to track the wave direction ($j=+1$ for forward-moving and $j=-1$ for backward-moving Alfv\'en waves with respect to the ambient magnetic field). Several studies have addressed the direction of propagation of Alfv\'enic turbulence \citep{bav03:alf}. In general it is expected that, closer to the Sun, most waves should propagate forward whereas at radial distances around 1~AU both wave intensities should be equal. In the current paper, therefore, we adopt the latter scenario.

\section{The Padian Code}

In the following we use a numerical Monte-Carlo code to compute the parallel diffusion coefficient of energetic particles for the turbulence model described above. A general description of the code and the underlying numerical techniques can be found elsewhere \citep{tau10:pad}. Specifically, the generation of turbulent magnetic fields corresponding to the composite model proceeds as follows: The slab and the two-dimensional components are generated separately using \citep{tau10:wav}
\be
\delta\f B(\f r,t)=\Re\;\sum_{n=1}^{N_m}\f e'_\perp A(k_n)\exp\!\left\{i\left[k_nz'+\beta_n-\omega(k_n)t\right]\right\},
\label{eq:dB}
\ee
where the wavenumbers $k_n$ are distributed logarithmically in the interval $k_{\text{min}} \leqslant k_n \leqslant k_{\text{max}}$ and where $\beta$ is a random phase angle. The important part is the time-dependence, which is taken as $\omega(k_n)=\;jv_{\mathrm A}(k_n)_{\parallel}$ for the slab modes (with $j=1$ for $n$ even and $j=-1$ for $n$ odd) and $\omega(k_n)=0$ for the two-dimensional modes. For the amplitude and the polarization vector, one has $A(k_n)\propto\sqrt{g(k_n)}$ and $\f e'_\perp\cdot\f e'_z=0$, respectively, with the primed coordinates determined by a rotation matrix with random angles so that $\f k\parallel\ez$ for slab modes and $\f k\perp\ez$ for 2D modes.

From the integration of the equation of motion, the parallel diffusion coefficient, $\kappa_\parallel$, is then calculated by averaging over an ensemble of particles and by determining the mean square displacement in the direction parallel to the background magnetic field as $\kappa_\parallel=(v/3)\lambda\pa=\langle(\varDelta z)^2\rangle/(2t)$.

For the simulation runs, the various parameters are summarized in Table~\ref{ta:param}. For the minimum and maximum wavenumbers included in the turbulence generator, the following considerations apply: (i) the \emph{resonance condition} states that there has to be a wavenumber $k$ so that $R_{\text L}k\approx1$, where $R_{\text L}$ denotes the particle's Larmor radius so that scattering predominantly occurs when a particle can interact with a wave mode over a full gyration cycle; (ii) the \emph{scaling condition} requires that $R_{\text L}\Om_{\text{rel}}t<L_{\text{max}}$, where $L_{\text{max}}\propto1/k_{\text{min}}$ is the maximum size of the system, which is given by the lowest wavenumber (for which one has $k_{\text{min}}=2\pi/\lambda_{\text{max}}$, thereby proving the argument). In practice, the second condition determines the minimum wavenumber while the first condition determines the maximum wavenumber.

Additionally, the relative strength of the turbulent magnetic field as compared to the homogenenous field is assumed to be $\delta B/B_0=0.5$ as suggested by \citet{ruf12:nlg}. Similar values were applied by \citet{heh12:mfp} for the interplanetary medium near Earth. Analytical estimates based on a WKB approach \citep{zan96:tur} indicate that the relative turbulence strength is somewhat less than unity.

Additional simulation parameters are chosen as $k_{\text{min}}\ell_{\mathrm{slab}}=10^{-4}$ and $k_{\text{max}}\ell_{\mathrm{slab}}=10^7$. Furthermore, the maximum simulation time can be expressed as $vt/\ell_{\mathrm{slab}}=10^2$ and the sum in Eq.~\eqref{eq:dB} extends over $N_m=2048$ wave modes.

An alternative method to simulate the transport of charged particles (so-called \emph{passive tracers}) by following the evolution of the magnetic field in a magnetohydrodynamic description \citep{mul07:tra,zan96:tur}. The comparison with either the Kolmogorov-based turbulence introduced above and with transport theories \citep{fum04:mhd} generally shows agreement; however, the dissipation range (see next section) might not be sufficiently resolved in MHD simulations \citep{ver12:mhd}.

\section{Comparison With Spacecraft Data}

\begin{figure}[tb]
\centering
\includegraphics[width=\linewidth]{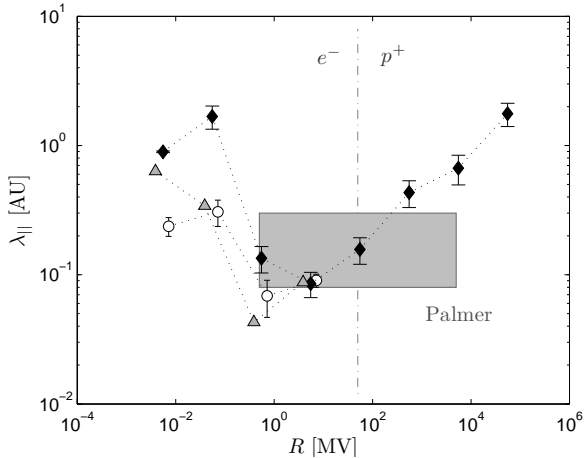}
\caption{The electron mean free path as a function of the particle rigidity for different dissipation properties. Whereas the black diamonds show the same case as in Fig.~\ref{ab:palmerPEb}, the other results use $k_d=3\times10^4\;(\mathrm{AU})^{-1}$ together with $p=3$ (gray triangles) and $k_d=3\times10^6\;(\mathrm{AU})^{-1}$ together with $p=3.5$ (white circles). For clarity, the horizontal positions are slightly shifted against each other.}
\label{ab:palmerPEcd}
\end{figure}

Analytically, particle scattering mean free paths in the interplanetary system can be obtained by fitting cosmic-ray observations to diffusion models. \citet{pal82:con} concluded that, for rigidities between $0.5$ and $5000$\,MV, the parallel mean free path is $0.08\,\mathrm{AU} \leqslant \lambda_{\parallel} \leqslant 0.3\,\mathrm{AU}$. This regime is usually visualized by the \emph{Palmer consensus range} \citep{bie94:pal}. In Figs.~\ref{ab:palmerPEb} and \ref{ab:palmerPEcd}, this box is compared to numerically calculated parallel mean free paths by using the best-fit parameters in the aforementioned \textsc{Padian} code. In Fig.~\ref{ab:palmerPEb}, the mean free paths are shown for electrons and for protons as the particle rigidity varies.

In Fig.~\ref{ab:palmerPEb}, the original measurements are shown in addition to the Palmer consensus range. Such is done because the latter was strongly criticized by \citet{rea99:acc} because it was obtained by excluding impulsive, scatter free events and Eastern solar events, thereby leading to values of the parallel mean free path smaller than the general case could be. Indeed, both the numerical and the experimental results shown in Fig.~\ref{ab:palmerPEb} correspond to mean free paths substantially larger than those of the Palmer consensus. It is interesting to notice that the numerical results compare well with the experimental data, even beyond the range of validity of Palmer consensus.

In Fig.~\ref{ab:palmerPEcd}, the electron mean free path is shown for varying dissipation range parameters $k_d$ and $p$. Thereby, the effects of the dissipation range and the dissipation spectral index are illustrated. For small rigidities, the mean free path is generally increased if the dissipation wavenumber, $k_d$, and/or the dissipation spectral range, $p$, are increased. Furthermore, Fig.~\ref{ab:palmerPEcd} shows that the electron mean free path can be as large as 2~AU, which is reminiscent of the works by \cite{per07:sup,per08:sup}, who inferred diverging mean free paths for electrons in the solar wind. The importance of alternative acceleration mechanisms without pitch-angle scattering is emphasized, which however cannot be reproduced in the simulations presented here because the magnetic turbulence model always leads to pitch-angle scattering for all rigidity values considered. Especially a superdiffusive mean free path \citep{tau10:sub} requires either purely two-dimensional turbulence \citep{sha08:two} or significantly stronger electric fluctuations \citep[e.\,g.,][]{tau10:wav}. Also, electron scatter free events are often described in the literature \citep[e.\,g.,][]{hag02:inj,lin05:sep}.

By comparing simulation results and observations, ones notes that the simulation data agrees with the bulk of the experimental events. In Fig.~1 of \citet{bie94:pal}, it can be seen that there are a number of electron events around 0.1\,MV with mean free paths of roughly $\lambda_\parallel\approx1$\,AU, a feature that is accurately reproduced by our simulation results. Likewise, there are proton events around $10^3$--$10^4$\,MV with $\lambda_\parallel\approx1$\,AU, which is equally well reproduced by the numerical results shown here. Therefore, the excellent agreement in comparison to the Palmer consensus range confirms the validity of the underlying model that has been constructed to describe the dynamics of the turbulent magnetic field. Furthermore, Fig.~\ref{ab:palmerR} illustrates the transport anisotropy, which is expressed by ratio of the perpendicular and parallel mean free paths. The average ratio is $\lambda_\perp/\lambda_\parallel\approx0.0379\pm0.0093$ for electrons and $\lambda_\perp/\lambda_\parallel\approx0.0597\pm0.0426$ for protons, in agreement with the Palmer values \citep{pal82:con,sha06:dyn}.

It should be noted that, in the numerical simulations, the effect of the turbulent electric field has been neglected intentionally in order to save computation time. Such is justified by the fact that, for shear Alfv\'en waves, the magnitude of the electric field is reduced by a factor of $v_{\mathrm A}/c\approx8.9\times10^{-3}$ as compared to the magnetic field amplitude \citep{rs:rays}. An additional simulation run confirmed that the inclusion of the turbulent electric field has a negligible effect on the spatial diffusion coefficients.

Furthermore, there has been a strong debate as to the form of the spectrum in the dissipation range, which could be either a second inertial range \citep{sah09:dis}  or a true dissipation range \citep{ale09:uni}. However, the modeling of the dissipation range in Eqs.~\eqref{specslab} and \eqref{spec2D} is purely heuristic and is therefore independent on the \emph{nature} of that spectral range. Instead, the simulation results are sensitive to the features of the ``dissipation'' range or, more precisely, on the parameters chosen for $k_d$ and $p$ so that new findings might alter the particle behavior. In the literature \citep[see Sec.~4.3 of][and references therein]{bie94:pal} it is argued that the spectral index must ultimately be larger than 3 for theoretical reasons but, at the same time, 3 provides a limit beyond which the mean square magnetic curl diverges. Therefore, the dissipation range spectral index was chosen as $p=3$ \citep[cf.][]{smi90:dis}.

\section{Summary and Conclusion}

A fundamental problem in cosmic-ray physics is the theoretical explanation of observed particle mean free paths in the solar system. \citet{pal82:con} described the conflict between observations and standard analytical theories. \citet{bie94:pal} revisited the problem by combining the quasi-linear transport theory with a more advanced model of interplanetary turbulence. This model takes account of spectral anisotropy, dissipation effects, and the turbulence dynamics. However, the applicability of quasi-linear theory used by these authors is questionable \citep{sha09:nli} and is, thus, not reliable.

An alternative approach to obtain scattering parameters theoretically is the application of test-particle codes. In such simulations there is no need for analytical transport theories. However, previous simulations have been performed for very simplified models of turbulence (e.\,g., static turbulence, no dissipation range in the spectrum). Therefore, no matter whether analytical or numerical descriptions of the transport had been used, the theory of particle diffusion has always been incomplete.

\begin{figure}[tb]
\centering
\includegraphics[width=\linewidth]{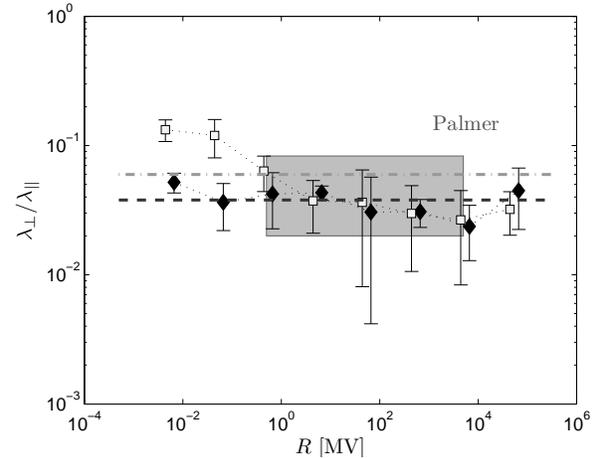}
\caption{The ratio of the perpendicular and parallel mean free path as a function of the particle rigidity. Shown are the results for electrons (black diamonds) and for protons (light squares) together with the respective error bars. The dashed and dot-dashed lines show the average values for electrons and protons, respectively. For comparison, the Palmer consensus range \citep{pal82:con,sha06:dyn} is shown. For clarity, the horizontal positions are slightly shifted against each other.}
\label{ab:palmerR}
\end{figure}

In the present paper we combined a realistic model for solar wind turbulence with the advanced \textsc{Padian} code. Numerical results for the parallel mean free path $\lambda_{\parallel}$ were compared with Palmer's consensus range in Figs.~\ref{ab:palmerPEb} and \ref{ab:palmerPEcd}. Furthermore, Fig.~\ref{ab:palmerR} shows the ratio of the perpendicular and parallel mean free paths, $\lambda_\perp/\lambda_\parallel$. According to the comparison shown here, solar wind observations of energetic particles can be reproduced with excellent accuracy. This positive result also confirms our present understanding of interplanetary turbulence and, accordingly, the slab/2D model.

It is important to note that all parameters were chosen according to standard interpretations of solar wind observations as, e.\,g., for: (i) the relative turbulence strength; (ii) the slab/2D ratio; (iii) the time-dependence of the slab component; (iv) the Alfv\'en velocity; (v) the dissipation range spectral index and wavenumber. Whereas a full parameter study was beyond the scope of the present investigation, instead we emphasize that the \citet{bie94:pal} observations of interplanetary particles were reproduced with excellent agreement by using test-particle simulations. Therefore, such theoretical results do no longer depend on uncertain analytical transport theories such as quasilinear theory. In light of these new results, it can be concluded that the development of analytical theories may still be based on the standard parameters and assumptions listed before, if: (i) the slab spectrum is chosen according to the new hybrid model, and (ii) a dissipation range is included.

\begin{acknowledgments}
A. Shalchi acknowledges support by the Natural Sciences and Engineering Research Council (NSERC) of Canada.
\end{acknowledgments}


\end{article}
\end{document}